\documentclass[smallextended,natbib]{article}
\usepackage{isolatin1}
\usepackage{graphicx}
\usepackage{algorithm}
\usepackage{algorithmic}
\usepackage{amssymb}
\usepackage{amsmath}
\usepackage{mathptmx}
\usepackage{rotating}
\usepackage{tikz}
\usepackage{natbib}

\newcommand{\N}{\ensuremath{\mathbb N}}
\newcommand{\R}{\ensuremath{\mathbb R}}

\DeclareMathOperator*{\argmin}{argmin}

\begin{document}

\title{Simple heuristics for the assembly line
  worker assignment and balancing problem}
\author{\small Mayron César O. Moreira, Alysson M. Costa\\
  \small Instituto de Ciências Matemáticas e de Computação, Universidade de São Paulo\\
  \small Av. Trabalhador são-carlense, 400, CP.~668, São Carlos -- SP, Brazil. CEP: 13560-970.\\
  \and \small Marcus Ritt\\
  \small Instituto de Informática - Universidade Federal do Rio Grande do Sul\\
  \and \small Antonio A. Chaves\\
  \small Faculdade de Eng. de Guaratinguetá - Universidade Estadual
  Paulista J.~de Mesquita Filho}
\date{}
\maketitle

\begin{abstract}
  We propose simple heuristics for the assembly line worker assignment
  and balancing problem. This problem typically occurs in assembly
  lines in sheltered work centers for the disabled. Different from the
  classical simple assembly line balancing problem, the task execution
  times vary according to the assigned worker. We develop a
  constructive heuristic framework based on task and worker priority
  rules defining the order in which the tasks and workers should be
  assigned to the workstations.  We present a number of such rules and
  compare their performance across three possible uses: as a
  stand-alone method, as an initial solution generator for
  meta-heuristics, and as a decoder for a hybrid genetic
  algorithm. Our results show that the heuristics are fast, they
  obtain good results as a stand-alone method and are efficient when
  used as a initial solution generator or as a solution decoder within
  more elaborate approaches.

\end{abstract}

\section{Introduction}

In assembly lines, products are assembled by means of the successive
execution of tasks in workstations. Each task has an execution time
and precedence relationships with other tasks. The \emph{simple
  assembly line balancing problem} (SALBP) concerns the decision of
allocating tasks to the workstations while respecting the partial
ordering of the tasks. Let $c$ be the cycle time of the line, i.e.,
the time spent at one of the workstations with the heaviest workload
and $m$ be the number of workstations. The problem is known as SALBP-1
when the goal is to minimize $m$ given a maximum allowed $c$, SALBP-2
when the goal is to minimize $c$ given a fixed $m$, SALBP-E when one
wants to minimize $m\cdot c$ and SALBP-F when the goal is to decide
whether a feasible solution does exist for given $m$ and $c$. For more
details on SALBP and many of its variants, we refer the reader to
\citet{baybars86survey}, \citet{scholl99balancing}
\citet{Scholl.Becker/2006}, \citet{boysen07classification}, and
\citet{boysen08assembly}.

The \emph{assembly line worker assignment and balancing problem}
(ALWABP) is an extension of the SALBP in which task execution times
are worker-dependent. This problem typically occurs when balancing
assembly lines with disabled workers, because a given worker might be
efficient on a certain subset of the tasks while being inefficient on
(or even unable of executing) other tasks. Analagous to the SALBP,
variants ALWABP-1, -2, -E and -F can also be defined.

The ALWABP has been introduced in the literature by
\citet{miralles07advantages} by means of a case study in an assembly
line of a Spanish sheltered work center for the disabled where a fixed
number of workers should be present in an assembly line whose
production rate should be maximized (ALWABP-2).  The same authors have
later developed a branch-and-bound algorithm for the problem, enabling
the solution of small-sized instances
\citep{miralles08branch}. Because of the problem complexity and the
need to solve larger instances, the literature has since then shifted
its efforts to heuristic methods. \citet{chaves09hybridb} have
proposed an elaborate clustering search algorithm that obtained good
solutions in reasonable computation times for instances of up to 19
workers. Their algorithm relies on a clustering procedure to group,
analyze and improve the solutions generated by a meta-heuristic
approach. In the search of a simpler method,
\citet{moreira09minimalist} have proposed a minimalist tabu search
algorithm based on the use of objective function penalties associated
to infeasible solutions to improve the algorithms ability to
effectively explore the search space. More recently,
\citet{blum11solving} have developed a beam search algorithm which has
been successful in improving the best known solutions obtained in the
previously mentioned studies.

In this study, we develop and test constructive heuristics for the
ALWABP-2. The heuristics generalize station-oriented assignment
procedures for the SALBP and apply them in search procedures for
finding the minimum cycle
time~\citep{scholl96simple}. Station-oriented assignment procedures
sequentially process the stations and assign tasks to the current
station in an order defined by a priority rule until it is maximally
loaded. The extension of such procedures to the ALWABP-2 is not
immediate, since the priority rules are often based on the task
execution times, a parameter that in the ALWABP depends on the worker
assigned to each workstation which is itself  an optimization decision.

The idea of our method is to use task and worker priority rules to
define which worker and which set of tasks will be assigned to each
workstation. We define 16 task priority rules and three worker
priority rules. The heuristic efficiency is compared by means of
computational tests to evaluate its performance both as a stand-alone
method as well as its ability to improve the convergence of more
sophisticated methods such as the aforementioned clustering search and
tabu search algorithms
\citep{chaves09hybridb,moreira09minimalist}. Moreover, the proposed
heuristic is also used as a solution decoder within a hybrid genetic
algorithm that optimizes explicit priorities for each task-worker
pair.

The remainder of this paper is organized as follows. Next, we present
a formal definition of the ALWABP and a mixed-integer linear model for
the problem. Then, the proposed heuristic is detailed in
Section~\ref{sec:proposed} and its use within a hybrid genetic
algorithm is described in Section~\ref{sec:biased}. A computational
study is presented in Section~\ref{sec:computationalstudy} and this
paper ends with some conclusions in Section~\ref{sec:conclusions}.

\section{Formal problem definition and mathematical model}
\label{sec:problem}

In this section we present a formal definition of the ALWABP-2. As
mentioned before, we denote by $c$ the cycle time and by $m$ the
number of workstations. Let $N$ be the set of tasks to be
allocated. The precedence constraints are given by a directed acyclic
graph $G=(N,E)$ over this set of tasks, where each edge $ij\in E$
indicates that task $i$ is an immediate predecessor of task $j$. We
also define an extended version of the precedence graph known as the
transitive closure of $G$, $G^*=(N,E^*)$, in which $ij \in E^*$ if
there is a path from $i$ to $j$ in $G$. In addition we will use the
following notation in the remainder of this paper:

\noindent\begin{tabular}{lp{0.65\linewidth}}
  $S$ & set of workstations;\\
  $W$ & set of workers, $|W| = |S|$; \\
  $t_i\in\N$ & time of task $i$ (SALBP);\\
  $t_{wi}\in\N\cup\{\infty\}$ & time of task $i$ when executed by
  worker $w$ (ALWABP);\\
  $I_w=\{ i\in N \mid t_{wi}=\infty\} $ & set of tasks infeasible for worker $w$;\\
  $P_i = \{ j\mid ji\in E\}$& set of immediate predecessors of task $i$;\\
  $P^*_i = \{ j\mid ji\in E^*\}$ & set of all predecessors of task $i$;\\
  $F_i = \{ j \mid ij\in E\}$ & set of immediate successors of task $i$;\\
  $F^*_i = \{ j \mid ij\in E^*\}$ & set of all successors of task $i$.
\end{tabular}

The goal of the problem is, given a fixed number of workstations,
$\overline m$, to find an assignment of tasks and workers to the
workstations minimizing $c$ and such that each task $i \in N$ is
assigned to a single station $s \in S$ and the precedence
relationships are respected, i.e., a task $i \in P_j$ can only be
assigned to the same workstation to which task $j$ has been assigned
or to workstations preceding it. We consider an equal number of
workstations and workers and, therefore, each worker $w \in W$ must be
assigned to a single station $s \in S$. Likewise, each workstation $s
\in S$ must receive a single worker $w \in W$. Workers cannot be
assigned independently from the task assignments, since the total load
of a workstation depends both on the tasks and the worker assigned to
it.

A mathematical model has been proposed by
\citet{miralles08branch}. The authors define binary variables
$x_{swi}$, equal to 1 only if task $i$ is assigned to worker $w$ at
workstation $s$, and binary variables $y_{sw}$, equal to 1 only if
worker $w$ is assigned to workstation $s$. The model can be written as
below:
\begin{align}
  \textbf{Minimize}\quad & c\label{1}\\
  \textbf{subject to}\quad & \sum_{w \in W} \sum_{s \in S} x_{swi} = 1 &&\forall i \in N, \label{2}\\
  &\sum_{s\in S} y_{sw} = 1 &&\forall w \in W, \label{3}\\
  &\sum_{w\in W} y_{sw} = 1 &&\forall s \in S, \label{4}\displaybreak[0]\\
  &\sum_{w\in W} \sum_{s\in S} s x_{swi} \leq \sum_{w\in W} \sum_{s\in S} s  x_{swj} &&\forall i,j \in N | i \in P_j, \label{5}\displaybreak[0]\\
  &\sum_{w\in W} \sum_{i\in N} t_{wi}  x_{swi} \leq c &&\forall s \in S, \label{6a}\displaybreak[0]\\
  &\sum_{i\in N} x_{swi} \leq |N| y_{sw} &&\forall w \in W, \forall s \in S, \label{7}\displaybreak[0]\\
  &x_{swi} = 0 &&\forall w \in W, \forall s \in S, \forall i \in I_w, \label{7a}\displaybreak[0]\\
  &y_{sw} \in \{0,1\} &&\forall s\in S, \forall w\in W, \label{8}\\
  &x_{swi} \in \{0,1\} &&\forall s\in S, \forall w\in W, \forall i\in N. \label{9}
\end{align}
Model~(\ref{1})--(\ref{9}) focuses on minimizing the cycle time for a
given number of workstations.  Constraints~(\ref{2}) guarantee that
each task is executed, and that it is done by a single worker, at a
single workstation.  Constraints~(\ref{3}) and~(\ref{4}) establish a
bijection between workers and workstations at a feasible solution,
i.e., every worker is assigned to a single workstation and \emph{vice
  versa}. Constraints~(\ref{5}) define the precedence
relations. Constraints~(\ref{6a}) establish that the cycle time is the
sum of the execution times of the tasks at the most charged
workstation, by forcing $c$ to be larger than or equal to the load
assigned to each station. Finally, constraints~(\ref{7}) indicate that
a task can only be assigned to a worker in a given workstation if that
worker is also assigned to the workstation.

The difficulty in solving this model has been reported in the
literature \citep{miralles08branch} and motivates the search for
efficient heuristic methods like the constructive heuristics presented
in the following section.

\section{Constructive heuristic methods}
\label{sec:proposed}

In this section we propose a constructive heuristic framework for the
ALWABP which is based on heuristics for the SALBP presented by
\citet{scholl96simple}. In order to simplify the presentation, we
describe the original heuristic in Section~\ref{subseccons} and
present the new developments in Section~\ref{subsecmod}.

\subsection{Constructive heuristics for the SALBP}
\label{subseccons}

The strategy used by \citet{scholl96simple} to solve the SALBP-2
relies on solving instances of SALBP-1 for different cycle times. A
given cycle time $c$ can be considered an upper bound for the SALBP-2
if the solution of the SALBP-1 needs, at most, the desired number of
workstations $\overline{m}$.

The solution of the SALBP-1 instances can rely upon exact or
approximate methods. Since the SALBP-1 is also an NP-hard problem,
heuristic methods are often used. These methods are usually based on
priority rules that order the tasks according to a given criterion and
assign them to the workstations accordingly. An important notion is
that of an \emph{available task}, which is an unassigned task whose
predecessors have already been assigned. Let $\overline{c}$ be the
desired cycle time and $t(S_k)$ the station time of workstation $k$. A
station-oriented procedure to solve the SALBP-1 can be easily
described in three steps:

\begin{algorithm}[!ht] \caption{Station-oriented assignment procedure for the SALBP-1}
  \small
  \label{alg:alg1}
  \begin{algorithmic}[1]
  \REQUIRE maximum cycle time $\overline{c}$
  \STATE Start at an empty workstation $k=0$, $t(S_k) = 0$. Set the
  unassigned tasks to $U = N$.
  \STATE Assign to the current workstation the available task $i$ with
  highest priority such that $t_i + t(S_k) \leq \overline{c}$ and set $U = U \backslash \{i\}$.
  \STATE If no more tasks are available, return $k$ (end). Otherwise,
  if there are available tasks, but none can be assigned to the
  current workstation $k$, set $k=k+1$, $t(S_k) = 0$. Go to step 2.
  \end{algorithmic}
\end{algorithm}

\citet{scholl96simple} mention the following selection of successful
priority rules, defining different orders with which the tasks should
be considered for assignment:

\begin{enumerate}
\item {\bf MaxF:} descending number of followers, $|F^*_i|$;
\item {\bf MaxIF:} descending number of immediate followers, $|F_i|$;
\item {\bf MaxTime:} descending task times, $t_i$;
\item {\bf MaxPW:} descending positional weights, $pw_i = t_i + \sum_{h\in F^*_i}  t_h$;
\item {\bf MaxTimeL:} descending task time divided by latest station, $t_i/L_i$;
\item {\bf MaxTimeSlack:} descending task time divided by slack, $t_i/s_i$
\end{enumerate}
where $E_i$ and $L_i$ are the earliest and latest workstations
(defined in Section~\ref{subsubsec:bounds}) to which a task can be
assigned and still obey the precedence constraints and the desired cycle
time $\overline{c}$, and where $s_i = L_i - E_i$ is the slack of task $i$
\citep{scholl99balancing}\footnote{If $E_i=L_i$, $s_i$ is set to some
  constant considerably smaller than one~\citep{scholl96simple}.}.

With these priority rules, an algorithm for the SALBP-2 can be
developed by searching for the smallest feasible
$\overline{c}$. \citet{scholl96simple} mention several search
procedures such as the lower bound method and binary search. In this
paper, we use the lower bound method in all our heuristics, since it
finds the smallest possible $\overline{c}$ for a given SALBP-1
heuristic.

\subsection{Extension to the ALWABP}
\label{subsecmod}

The extension of the SALBP priority rules to the ALWABP faces two main
difficulties. First, most of the priority rules rely on task execution
times, which are not well defined for the case of the ALWABP since
they depend on the designated worker. Second, there is no strategy to
select the worker to be assigned to each workstation. In the
following, we propose modifications to the original algorithm in order
to cope with these two difficulties.

\subsubsection{Task priority rules}

From the six priority rules suggested by \citet{scholl96simple}, the
first two can be used unchanged in ALWABP, because they do not depend
on task execution times. Rules MaxTime and MaxPW do depend on task
execution times but can still be adapted if one considers, for
instance, minimum, maximum or averaged values of these times over all
workers. In particular, for the original rule MaxTime based on task
execution times, we considered ascending and descending orderings for
each possibility. Finally, the last two rules depend on the earliest
and latest stations to which a task can be assigned; we do decided to
not apply these rules since the corresponding station bounds are
rather weak for ALWABP.

The proposed adaptions lead to eleven rules for the ALWABP derived
from existing rules for the SALBP. We further propose five new rules
that try to reflect the structure of the ALWABP by using the original
task execution times per worker. The idea is to prioritize tasks when
workers that quickly execute these tasks are being considered. The
full set of priority rules are listed and described in the following,
where $\overline{w}_i \in\mathrm{\argmin}_{w \in W} t_{wi}$ is some
fastest worker in the execution of task $i$ and $t_i^- =
\textrm{min}_{w\in W} t_{wi}$, $t_i^+ = \textrm{max}_{w\in W} t_{wi}$,
$\overline{t}_i = \sum_{w\in W} t_{wi}/|W|$ are the minimum, maximum
and mean execution time of task $i$, respectively. If task $i$ is
infeasible for worker $w$, i.e., $t_{wi}= \infty$, we set $t_{wi} =
\overline{c}$ for the purposes of computing $t_i^+$ and
$\overline{t}_i$, where $\overline{c}$ is cycle time under
consideration (see Algorithm~\ref{alg:alg2}).

\begin{enumerate}
\item {\bf MaxF}: descending number of followers, $|F^*_i|$;
\item {\bf MaxIF}: descending number of immediate followers, $|F_i|$;
\item {\bf MaxTime$^-$}: descending minimum task times, $t_i^-$;
\item {\bf MaxTime$^+$}: descending maximum task times, $t_i^+$;
\item {\bf Max$\overline{\textrm{Time}}$}: descending average task times, $\overline{t}_i$;
\item {\bf MinTime$^-$}: ascending minimum task time, $t_i^-$;
\item {\bf MinTime$^+$}: ascending maximum task time, $t_i^+$;
\item {\bf Min$\overline{\textrm{Time}}$}: ascending average task time, $\overline{t}_i$;
\item {\bf MaxPW$^-$}: descending minimum positional weights, $pw_i^- = t_i^- + \sum_{h\in F^*_i} t_h^-$;
\item {\bf MaxPW$^+$}: descending maximum positional weights, $pw_i^+ = t_i^+ + \sum_{h\in F^*_i} t_h^+$;
\item {\bf Max$\overline{\textrm{PW}}$}: descending average positional weights, $\overline{pw_i} = \overline{t}_i + \sum_{h\in F^*_i} \overline{t}_h$;
\item {\bf MinD(w)}: ascending difference to best worker, $t_{wi} - t_{\overline{w}_ii}$;
\item {\bf MinR(w)}: ascending ratio to best worker, $t_{wi}/t_{\overline{w}_ii}$;
\item {\bf MaxFTime(w)}: descending number of followers per time, $|F_i|/t_{wi}$;
\item {\bf MaxIFTime(w)}: descending number of immediate followers per time, $|F^*_i|/t_{wi}$;
\item \textbf{MinRank(w)}: ascending rank of worker's execution time, $|\{ w'\in W \mid t_{w'i}<t_{wi}\}|$.
\end{enumerate}

Rules MinD(w) and MinR(w) prioritize tasks for which the current
worker w is faster than other workers while MaxFTime(w) and
MaxIFTime(w) balance two desired characteristics of a task: it should
be executed quickly by the current worker $w$ and it should make the
largest number of tasks available. For the current worker $w$, rule
MinRank(w) gives preference to a task with a short execution time
compared to other workers, but does not depend on concrete execution
times. The other rules follow the original rationale.

\subsubsection{Worker priority rules}

Many of the task priority rules proposed above analyze the efficiency
of a given worker for the task in comparison to the other
workers. This suggests that Algorithm~\ref{alg:alg1} needs to be
modified to include a worker selection loop in which all possible
workers are tested and, according to a worker selection criterion, one
is chosen for the current workstation. This modification is reflected
in Algorithm~\ref{alg:alg2}. We also add an outer \emph{while} loop on
cycle time tentative values, in order to present a self-contained
algorithm for the ALWABP-2.

\begin{algorithm}[!ht] \caption{Station-oriented assignment procedure for the ALWABP-2}
  \small
  \label{alg:alg2}
  \begin{algorithmic}[1]
  \STATE $\overline{c}$ = lower bound on the cycle time.
  \WHILE{solution not found}
  	\STATE Start at an empty workstation $k=0$, $t(S_k) = 0$.
  	\STATE Set the unassigned workers to $U_w = W$, set the unassigned tasks to $U = N$.
  	\FOR{each worker $w$ in $U_w$}
  	\STATE Compute $T_w$, the set of tasks that would be assigned to worker $w$ according to the task priority rule used.
  	\ENDFOR
  	\STATE Assign the worker $w$ with the highest priority and its tasks
  	$T_w$ to the station $k$. $U_w = U_w \backslash \{w\}$, $U = U\backslash T_w$, $k=k+1$.
  	\IF{ $U_w \neq \emptyset $ }
    	\STATE Go to step 4.
  	\ELSE
    \IF{$U = \emptyset$}
        \RETURN $\overline{c}$
    \ELSE
        \STATE $\overline{c} = \overline{c} + 1$ 
    \ENDIF
  	\ENDIF
  \ENDWHILE
  \end{algorithmic}
\end{algorithm}

Algorithm~\ref{alg:alg2} tests, for each workstation, all unassigned
workers. In step 6, the algorithm computes the tasks that can be
assigned to the current station if the worker under consideration is
also assigned to that station. Note that these tasks depend on the
priority rule, on the worker and on the tentative cycle time being
considered. In step 8, according to the selected tasks, it chooses the
worker which will be assigned to the current workstation. Let
$t_i^{-}(W)=\min_{w\in W} t_{wi}$ be the minimum execution time of
task $i$ among workers $W$. We propose three worker selection rules:

\begin{itemize}
\item {\bf MaxTasks}: descending number of assignable tasks, $|T_w|$;
\item {\bf MinBWA}: ascending best worker assignment cycle
  MinBWA($U\setminus T_w$,$U_w\setminus\{w\}$) (Algorithm~\ref{alg:algminbwa}).
\item {\bf MinRLB}: ascending restricted lower bound, 
$$\frac{\sum_{i \in U\backslash T_w} \; t_i^{-}(U_w\setminus\{w\})}{|U_w| -1}\mathrm{.}$$
\end{itemize}

\begin{algorithm}[!ht] \caption{Best worker assignment cycle (MinBWA)}
  \small
  \label{alg:algminbwa}
  \begin{algorithmic}[1]
  \STATE \textbf{input:} a set of tasks $T$ and workers $W$
  \STATE Set $t_w=0$ for all $w$ in $W$
  \FOR{each task $i$ in $T$}
  \STATE Determine the set of workers executing $i$ fastest,
$W':=\argmin_{w\in W} t_{wi}$
  \STATE Select a worker $w'\in W'$ of minimum load $t_{w'}=\min_{w\in W'}
t_w$
  \STATE $t_{w'}=t_{w'}+t_{w'i}$
  \ENDFOR
  \RETURN the maximum cycle time $c=\max_{w\in W} t_w$
  \end{algorithmic}
\end{algorithm}

The first rule gives preference to the worker that was able to execute
the largest number of tasks. The last two criteria compute estimates
on the cycle time for the problem considering the yet unassigned
workers and tasks, and give preference to the worker that reduces this
estimate. They differ on the way the estimates are obtained. Rule
MinBWA computes the cycle time that would be obtained if each
remaining worker were assigned the tasks which he executes fastest (in
case of ties, the task is attributed to the currently less charged
worker). Rule MinRLB computes a lower bound for the cycle time on the
remaining workstations by dividing the sum of the minimum execution
times of the remaining tasks by the number of remaining workers.

Each of these three   worker priority rules can be combined with one of
the above $16$   task priority rules. Moreover, if one considers that
the station-oriented assignment can be run in a forward (considering
the original precedence graph) or backward (considering the reversed
precedence graph) manner, a total of $96$   possible combinations are
possible. The efficiency of these combinations is compared in
Section~\ref{sec:computationalstudy}.

\subsubsection{Bounds on the minimum cycle time}
\label{subsubsec:bounds}

In order to reduce the number of calls made to
Algorithm~\ref{alg:alg2} and, therefore, improve the efficiency of the
method, one might investigate good lower bounds for the
ALWABP-2. Valid lower bounds for the SALBP-2 can be used by
substituting the task execution time in the original bound by the
minimum task execution time over all workers $t_i^-$. With this
strategy, we are able to use lower bounds LC1, LC2, and LC3 from
\citet{Scholl.Becker/2006} in our search procedures. LC1 and LC2 are
defined as
\begin{align*}
  \mathrm{LC1} & := \max \biggl\{ \max_{i\in N} t_i^-, \biggl\lceil\sum_{i\in N} t_i^- / m\biggl\rceil \biggr\}\\
  \mathrm{LC2} & := \max \biggl\{ \sum_{0\leq i\leq k} t^-_{km+1-i} \mid 1\leq k\leq \bigl\lfloor (n-1)/m \bigr\rfloor \biggr\}\mathrm{.}
\end{align*}

The bound LC3 is obtained by \emph{destructive
  improvement}~\citep{Klein.Scholl/1999} using bounds on the earliest
station $E_i(c)$ and latest station $L_i(c)$ for each task $i$, and is
defined as the smallest $c$ greater than or equal to an initial lower
bound $\underline{c}$ such that $E_i(c)\leq L_i(c)$ for all tasks
$i$. When using minimum task times, the station bounds are defined as
\begin{align*}
  E_i(c) & := \biggl\lceil \biggl(\sum_{j\in P^*_i} t^-_j + t^-_i\biggr)/c\biggr\rceil\\
  L_i(c) & := m+1 - \biggl\lceil \biggl(t^-_i + \sum_{j\in F^*_i} t^-_j\biggr) / c\biggr\rceil\mathrm{.}
\end{align*}

For a given cycle time $c$, these bounds can be improved by
preprocessing the ALWABP-2 instance as follows. For tasks $i,k$ let
$O(i,k)=(F_i^*\cap P_k^*) \cup (F_k^*\cap P_i^*)$ be the tasks that
must be executed after $i$ and before $k$ or {\it vice-versa}, and let
$R(i,k)=O(i,k)\cup\{i,k\}$. If some task $i$ can only be executed by a
given worker $w$, then another task $k$ can be executed by the same
worker $w$ only if the total execution time of all tasks between $i$
and $k$ does not exceed $c$, i.e., $\sum_{j\in R(i,k)} t_{wj}\leq
c$. Otherwise, task $k$ is infeasible for worker $w$. In this latter
case, a reduced instance for a cycle time $c$ can be obtained by
setting the execution time $t_{wk}$ of all such tasks to $\infty$,
making them infeasible for worker $w$.

As a fourth bound we use the linear relaxation of model
\eqref{1}--\eqref{9}. For SALBP-2 the linear relaxation of the
ILP-model is known to be weak, but this bound proves to be useful for
ALWABP-2, because the use of minimum execution times weakens the lower
bounds LC1, LC2, and LC3. Since this bound is the most expensive
computationally, it is used only in the genetic algorithm presented in
the next section.

\section{A hybrid genetic algorithm for ALWABP-2}
\label{sec:biased}

In this section we propose a hybrid biased random-key genetic
algorithm (HGA) based on the constructive heuristic of
Section~\ref{sec:proposed}. Random-key genetic algorithms are
particularly useful for sequencing problems, and have been proposed by
\citet{Goncalves.Almeida/2004} for the SALBP-1 and, by
\citet{Mendes.etal/2009} for resource constrained project scheduling.

A genetic algorithm~\citep{Goldberg/1989,Holland.1975} is a
population-based meta-heuristic. Each individual in the population has
a \emph{chromosome} which codifies a solution to a problem instance. A
chromosome consists of a collection of \emph{genes} each of which can
take a value among several \emph{alleles}. A population generates
offspring by crossover between two individuals and mutation. The
probability of an individual to participate in a crossover is
proportional to its fitness, making it more likely that fitter
individuals pass genetic material to the next generation. A selection
rule determines which individuals from the current generation and the
offspring form the next generation.

\citet{Bean/1994} introduced random-key genetic algorithms (RKGA). In
a RKGA the chromosome is a vector in $\R^n$ (where $n$ is a
problem-dependent parameter) and a \emph{decoder} maps it to a
feasible solution. For example, in problems of sequencing jobs, the
order of the genes' values defines a permutation of the jobs. The
advantage of an RKGA is that the decoding process guarantees that each
chromosome corresponds to a feasible solution and that the search can
be done in a problem-independent way in the space $\R^n$.

The evolutionary dynamics of a RKGA is as follows. All $p$ individuals
in the current population are sorted by their fitness value, and the
best $p_e$ \emph{elite} individuals are copied into the next
generation. A RKGA replaces mutation by \emph{immigration} of a small
percentage $p_r$ of random individuals into the next generation. The
remaining $p-p_e-p_r$ individuals are offspring. To generate
offspring, two individuals are selected at random from the current
population and are combined by \emph{uniform crossover}. In uniform
crossover, the offspring receives each allele independently with
probability $q\geq 0.5$ from the first parent and probability $1-q$
from the second parent.

A biased random-key genetic algorithm differs from a RKGA in
the way it selects parents for crossover. The first parent is drawn
randomly from the elite set, and the second one randomly from the
remaining individuals. In this way, the offspring inherits with a
higher probability an allele from an elite parent. A hybrid genetic
algorithm applies a local search method to its individuals to improve
their fitness.

To construct a HGA for the ALWABP-2, we substitute the task priority
rules presented in Section 3 by explicit priorities. Instead of using
a rule to define a priority $\mathrm{pr}(w,i)$ for each worker $w$ and
task $i$, the chromosome of an individual in the HGA is a matrix
$p_{wi}$ of priorities in $[0,1]$. The fitness of an individual is the
result of Algorithm 2 run with these priorities.
Algorithm~\ref{alg:alg2} uses the bounds described in
Section~\ref{subsubsec:bounds}, including the linear relaxation lower
bound, when searching for a feasible cycle time. To find a feasible
solution, it tries to allocate all tasks in a forward as well as a
backward manner.

After decoding, a local search is applied to improve the solution. The
local search uses two types of well-known
moves~\citep{Scholl.Becker/2006}: a \emph{shift} of a task from one
station to another, and a \emph{swap} of two tasks between different
stations. We allow also a sequence of two shifts, where the first one
does not improve the solution. A third move type applied in the local
search is a \emph{swap of two workers} between two stations. All moves
are applied only when they produce a feasible solution and are able to
reduce the number of stations with load equal to the cycle time.

The fitness of an individual is a pair $(c,l)$, where $c$ is the
smallest cycle time found after local search, and $l$ is the
normalized total station time, i.e., the sum of all task execution
times divided by $mc$. Fitness is ordered lexicographically. The
second component allows the HGA to reduce, for a constant cycle time
$c$, the station times of a solution.

The initial population is seeded with individuals whose priorities are
set according to the $16$ proposed task priority rules. If the
population size is less than $16$ we select the $p$ best
individuals. For populations larger than $16$, the remaining
individuals of the population are initialized randomly. The
pseudo-code for the HGA is shown in Algorithm~\ref{alg:brkga}.

\begin{algorithm}[!ht] \caption{HGA for the ALWABP-2}
  \label{alg:brkga}
  \begin{algorithmic}[1]
    \STATE \textbf{input:} population sizes $p$, $p_e$, $p_r$, crossover probability $q$;
    \STATE\COMMENT{We denote by $f(i)$ the fitness of individual $i$}
    \STATE $P:=\{(i_1,f(i_1)),\ldots,(i_{16},f(i_{16}))\}$ where $i_k\in[0,1]^{|W|\times|N|}$, $1\leq k\leq 16$ is set using task priority rule $k$;
    \STATE $P:=P\cup\{(i_{17},f(i_{17})),\ldots,(i_p,f(i_p))\}$ where each $i_k$, $17\leq k\leq p$ is a random matrix in $[0,1]^{|W|\times|N|}$;
    \STATE sort $P$ by increasing cycle times $f(i_k)$ and keep the $p$ best individuals;
    \WHILE {stopping criterion is not satisfied}
      \STATE $P':=\{(i_1,f(i_1)),\ldots,(i_{p_e},f(i_{p_e})\}$;   \COMMENT{ elite passes to the next generation }
      \FOR {$k=1,\ldots,p-p_e-p_r$} 
        \STATE select a random individual $i$ from $\{i_1,\ldots,i_{p_e}\}$;
        \STATE select a random individual $j$ from $\{i_{p_e+1},\ldots,i_{p}\}$;
        \STATE $o:=\mathrm{crossover}(i,j,q)$;
        \STATE $\mathrm{localsearch}(o)$;
        \STATE $P':=P'\cup\{(o,f(o))\}$;
      \ENDFOR
      \FOR {$k=1,\ldots,p_r$}  
        \STATE $P':=P'\cup\{(i,f(i))\}$ where $i$ is a random matrix in $[0,1]^{|W|\times|N|}$;
      \ENDFOR
      \STATE $P:=P'$;
      \STATE sort $P$ by increasing cycle times $f(i_k)$;
    \ENDWHILE
    \RETURN best individual $i_1$;
\end{algorithmic}
\end{algorithm}

\section{Computational study}
\label{sec:computationalstudy}

In this section, we present a computational study of the proposed
lower bounds, the constructive heuristics, and the hybrid genetic
algorithm. In all experiments, we have used the instances available in
the literature~\citep{chaves07clustering}\footnote{The instances can
  be found at
  \texttt{http://www.feg.unesp.br/\~{}chaves/Arquivos/ALWABP.zip}.}.
They are grouped in four families: Roszieg, Heskia, Wee-Mag and Tonge,
each one containing $80$ instances. The instances were generated from
the corresponding SALBP instances such that they contain $10$
instances for each combination of five experimental factors at a low
and a high level. The factors are the number of tasks, the number of
workers, the order strength\footnote{The order strength is defined as
  the fraction of present precedence relations compared to the maximum
  possible, i.e., $2|E^*|/(n(n-1))$.}, the variability of the task
execution time, and the number of infeasible task-worker pairs. For
instances with low variability, the task execution times are drawn
uniformly from the interval $[1,t_i]$, where $t_i$ is the task
execution time as defined by the SALBP instance. When the variability
is high, this interval was $[1,3t_i]$. The number of infeasible
task-worker pairs is $10\%$ and $20\%$ on the low and the high level,
respectively. The main characteristics for each group of instances
(number of tasks, number of workers and the order strength of the
precedence network) are listed in Table~\ref{tab:1}.

\begin{table}[h!]
\caption{Instance characteristics.}\label{tab:1}
\begin{tabular}{ccccc}
\hline\noalign{\smallskip}
\textbf{Family} &   \mbox{\boldmath${|N|}$} & \mbox{\boldmath${|W|}$}
&   \textbf{Order Strength} & \textbf{BKV/LB (\%)}\\
\noalign{\smallskip}\hline\noalign{\smallskip}
Roszieg & 25 & 4 (groups 1-4) or 6 (groups 5-8) &  71.67 & 16.75 \\
Heskia & 28 & 4 (groups 1-4) or 7 (groups 5-8) &  22.49 & 57.91 \\
Tonge & 70 & 10 (groups 1-4) or 17 (groups 5-8) &  59.42 & 57.91 \\
Wee-Mag & 75 & 11 (groups 1-4) or 19 (groups 5-8) & 22.67 & 76.70 \\
\noalign{\smallskip}\hline
\end{tabular}
\end{table}

Our results are given as relative deviations over the objective value
of the optimal solution or the best known value. For instance families
Heskia and Roszieg the optimal solutions are known (for a complete
table see~\citet{blum11solving}). For instance families Wee-Mag and
Tonge we compare with the best known value defined as the minimum of
the best values found by a clustering search, a tabu search and an
iterated beam search as published in the
literature~\citep{chaves09hybridb,moreira09minimalist,blum11solving}. All
computation times reported are in seconds of real time.

To be able to estimate the quality of the lower bounds and the best
known values, we report in the last column of Table~\ref{tab:1} the
average relative deviation of the optimal solution or best known value
from the lower bound. In comparison over all $320$ instances lower
bound LC1 was maximum in $118$ cases, LC2 in none, LC3 in $162$ cases
and the linear relaxation of the ILP model \eqref{1}--\eqref{9} in
$235$ cases.

\subsection{Constructive heuristic methods for ALWABP-2}

For the SALBP, it is known that breaking ties between tasks of the
same priority can have a significant effect on the solution
quality~\citep{Talbot.etal/1986}. Therefore we tested several
tie-breaking rules for both task and worker selection. For selecting
tasks, we used as a first-level tie-breaker the descending number of
immediate followers (MaxIF) and as a second-level tie-breaker the task
execution time for the current worker ($t_{wi}$). For the worker
selection, as a first-level tie-breaker, we used MinRLB for rules
MaxTask and MinBWA and MaxTask for rule MinRLB.  As a second-level
tie-breaker, we used the workstation idle time. Both for tasks and
workers, the original index of the task or worker was considered as a
third-level tie-breaker, in order to make the heuristics
deterministic.

Another important implementation aspect is the fact that some of the
priority rules can be applied statically or dynamically with respect
to the worker selection loop. For instance, task priority rules
MinD(w) and MinR(w) as well as all task priority rules that rely on
the calculation of a minimum, maximum or average value of a parameter
over all workers are prone to changes as the set of available workers
is reduced. Preliminary tests showed that the dynamic strategy using
only the available workers presented better results and was used in
all subsequent tests. For the sake of simplicity and due to the
computational efficiency of Algorithm 2, we used LC1 to obtain the
lower bound on the cycle time in line 1 of Algorithm~2.

Under the considerations above, all combinations of task and worker
priority rules were considered. The heuristics were implemented in
\emph{C}, using the compiler gcc 4.4, under the Linux Ubuntu operating
system. For the tests we used a PC with a Core 2 Duo $2.2$ GHz
processor and $3$ GB of main memory.

The first conclusion that could be drawn was that for all possible
task selection rules, worker selection rule MinRLB consistently
obtained better results when compared to rules MaxTasks and MinBWA,
which were, therefore, discarded. The results obtained for the 16 task
priority rules, worker selection rule MinRLB and strategies backward
and forward are presented in Table~\ref{tab:resultadosHeuristicas}. In
the table, we present the average and maximum deviation of the
heuristic solution with respect to the best known values and the
average and maximum computation time.

\begin{sidewaystable}[htbp]
\footnotesize
\caption{Results for the 16 task priority rules, worker selection rule MinRLB and strategies backward and forward over 320 instances.}
\begin{tabular}{llrrrrrrrr}
\hline
Allocation & \multicolumn{1}{l}{} & \multicolumn{1}{l}{} & \multicolumn{1}{l}{} & \multicolumn{1}{l}{} & \multicolumn{1}{l}{} & \multicolumn{1}{l}{} & \multicolumn{1}{l}{} & \multicolumn{1}{l}{} \\
\hline\noalign{\smallskip}
 & Task rule & \multicolumn{1}{l}{MaxF} & \multicolumn{1}{l}{MaxIF} & \multicolumn{1}{l}{MaxTime$^-$} & \multicolumn{1}{l}{MaxTime$^+$} & \multicolumn{1}{l}{Max$\overline{\text{Time}}$} & \multicolumn{1}{l}{MinTime$^-$} & \multicolumn{1}{l}{MinTime$^+$} & \multicolumn{1}{l}{Min$\overline{\text{Time}}$} \\
\hline
Forward & av. dev. (\%) & 25.5\% & 30.8\% & 26.6\% & 30.9\% & 28.0\% & 55.5\% & 49.4\% & 53.4\% \\
 &  max. dev. (\%) & 108.0\% & 145.5\% & 132.0\% & 132.0\% & 108.0\% & 158.5\% & 150.0\% & 140.0\% \\
 &  av. time & 0.02 & 0.02 & 0.02 & 0.02 & 0.02 & 0.02 & 0.02 & 0.02 \\
 &  max. time & 0.08 & 0.08 & 0.08 & 0.10 & 0.10 & 0.10 & 0.13 & 0.10 \\
\cline{2-10}\noalign{\smallskip}
 & Task rule & \multicolumn{1}{l}{MaxPW$^-$} & \multicolumn{1}{l}{MaxPW$^+$} & \multicolumn{1}{l}{Max$\overline{\text{PW}}$} & \multicolumn{1}{l}{MinD(w)} & \multicolumn{1}{l}{MinR(w)} & \multicolumn{1}{l}{MaxFTime(w)} & \multicolumn{1}{l}{MaxIFTime(w)} & \multicolumn{1}{l}{MinRank(w)} \\ \cline{2-10}
 &  av. dev. (\%) & 17.0\% & 20.2\% & 19.0\% & 27.2\% & 25.6\% & 28.5\% & 30.8\% & 23.4\% \\
 &  max. dev. (\%) & 108.0\% & 108.0\% & 108.0\% & 136.0\% & 136.0\% & 132.0\% & 145.5\% & 112.0\% \\
 &  av. time & 0.04 & 0.04 & 0.03 & 0.02 & 0.02 & 0.02 & 0.02 & 0.03 \\
 &  max. time & 0.26 & 0.19 & 0.17 & 0.09 & 0.10 & 0.11 & 0.10 & 0.17 \\
\hline\noalign{\smallskip}
& Task rule & \multicolumn{1}{l}{MaxF} & \multicolumn{1}{l}{MaxIF} & \multicolumn{1}{l}{MaxTime$^-$} & \multicolumn{1}{l}{MaxTime$^+$} & \multicolumn{1}{l}{Max$\overline{\text{Time}}$} & \multicolumn{1}{l}{MinTime$^-$} & \multicolumn{1}{l}{MinTime$^+$} & \multicolumn{1}{l}{Min$\overline{\text{Time}}$} \\
\cline{2-10}
Backward & av. dev. (\%) & 33.0\% & 39.7\% & 32.2\% & 34.7\% & 31.0\% & 53.9\% & 49.9\% & 55.2\% \\
 & max. dev. (\%) & 218.2\% & 372.7\% & 177.8\% & 177.8\% & 125.9\% & 203.7\% & 272.7\% & 203.7\% \\
 & av. time & 0.02 & 0.02 & 0.02 & 0.02 & 0.02 & 0.03 & 0.03 & 0.03 \\
 & max. time & 0.09 & 0.09 & 0.11 & 0.10 & 0.10 & 0.11 & 0.11 & 0.15 \\
\cline{2-10}\noalign{\smallskip}
 & Task rule & \multicolumn{1}{l}{MaxPW$^-$} & \multicolumn{1}{l}{MaxPW$^+$} & \multicolumn{1}{l}{Max$\overline{\text{PW}}$} & \multicolumn{1}{l}{MinD(w)} & \multicolumn{1}{l}{MinR(w)} & \multicolumn{1}{l}{MaxFTime(w)} & \multicolumn{1}{l}{MaxIFTime(w)} & \multicolumn{1}{l}{MinRank(w)} \\ \cline{2-10}
 & av. dev. (\%) & 26.2\% & 28.7\% & 27.8\% & 31.8\% & 29.2\% & 36.9\% & 40.0\% & 28.0\% \\
 & max. dev. (\%) & 218.2\% & 213.6\% & 200.0\% & 372.7\% & 177.8\% & 372.7\% & 372.7\% & 372.7\% \\
 & av. time & 0.06 & 0.05 & 0.04 & 0.02 & 0.02 & 0.02 & 0.02 & 0.03 \\
 & max. time & 0.27 & 0.23 & 0.23 & 0.11 & 0.10 & 0.10 & 0.12 & 0.15 \\
\hline
Best over all rules &  av. dev. (\%) & 9.6\% & \multicolumn{1}{l}{} & \multicolumn{1}{l}{} & \multicolumn{1}{l}{} & \multicolumn{1}{l}{} & \multicolumn{1}{l}{} & \multicolumn{1}{l}{} & \multicolumn{1}{l}{} \\
 & max. dev. (\%) & 55.6\% & \multicolumn{1}{l}{} & \multicolumn{1}{l}{} & \multicolumn{1}{l}{} & \multicolumn{1}{l}{} & \multicolumn{1}{l}{} & \multicolumn{1}{l}{} & \multicolumn{1}{l}{} \\
 & av. time & 0.83 & \multicolumn{1}{l}{} & \multicolumn{1}{l}{} & \multicolumn{1}{l}{} & \multicolumn{1}{l}{} & \multicolumn{1}{l}{} & \multicolumn{1}{l}{} & \multicolumn{1}{l}{} \\
 & max. time & 3.66 & \multicolumn{1}{l}{} & \multicolumn{1}{l}{} & \multicolumn{1}{l}{} & \multicolumn{1}{l}{} & \multicolumn{1}{l}{} & \multicolumn{1}{l}{} & \multicolumn{1}{l}{} \\
\hline
\end{tabular}
\label{tab:resultadosHeuristicas}
\end{sidewaystable}

From the results in Table~\ref{tab:resultadosHeuristicas}, we can
conclude that the heuristics can obtain feasible solutions for the
ALWABP in very small computation times (not more than $0.27$ seconds
in the worst case).  The quality of the obtained solution is sensitive
to the task priority rule used, with the rules based on the positional
weights presenting the best results. Indeed, criterion MaxPW$^-$
yielded average deviations of $17.0$ percent and $26.2$ percent for
the forward and backward searches, respectively. The diversity of the
task priority rules can be evaluated by considering, for each
instance, the best obtained solution over all rules. This computation
yields an average deviation of $9.6$ percent in less than one second
of average computation time. The application of the local search
described in Section~\ref{sec:biased} presented only marginal gains.

These results are competitive with those obtained by the clustering
search algorithm of \citet{chaves09hybridb} which obtained average
deviation values of 21.1 percent and with the tabu search of
\citet{moreira09minimalist} which obtained average results of 26.8
percent, in larger computation times. Indeed, both meta-heuristic
methods can benefit from an initial solution such as those obtained
with our proposed strategy. We ran both methods initialized with the
solutions obtained with criterion MaxPW$^-$ and forward
allocation. The clustering search algorithm and the tabu search
algorithm then presented average deviations of $11.7$ percent and
$9.9$ percent respectively. These results can be further improved if
the best solution (over all priority criteria) is used as initial
solution. In this case, the clustering search and the tabu search
yielded average results of $7.8$ percent and $7.4$ percent,
respectively (see Table~\ref{tab:cmp}). These results suggest the
advantage of quickly obtaining good initial solutions such as those
presented in this article.

\subsection{A HGA for ALWABP-2}

Based on the results of the previous section we chose MinRLB as the
worker priority rule for the HGA. In a preliminary experiment on three
selected instances of all four families, we tested the performance of
the HGA on populations of size $P=40,70,100$, and for uniform
crossover probabilities of $p=0.5$, $0.6$, $0.7$, $0.8$, $0.9$. Based
on this experiment, we chose a population size of $P=100$ and a
crossover probability of $p=0.5$. We ran the HGA with these parameters
$20$ times with different random seeds on all $320$ instances. The
execution stopped after $200$ iterations, or $100$ iterations without
improvement of the incumbent. We used a PC with a $2.8$ GHz Intel Core
i7 930 processor with $3$ GB of main memory for the experiments.

Table~\ref{tab:cmp} compares the results of the HGA with the
clustering search (CS) of \citet{chaves09hybridb}, the tabu search
(TS) of \citet{moreira09minimalist}, and the iterated beam search
(IBS) of \citet{blum11solving}. For each group of the four instance
families, and for each of the four meta-heuristics, we report the
relative deviation of the best value found from the best known value,
the relative deviation of the average value from the best known value,
the average total computation time, and the average computation time
to find the best solution. We report only the solution value for TS,
since the method is deterministic, and only the time to find the best
value for IBS, since \citet{blum11solving} do not report total
computation time.  The results of CS and of the TS have been obtained
using the best constructive heuristic to generate the initial
solution.

\begin{sidewaystable}[htbp]
\caption{Comparison of four meta-heuristics for the ALWABP-2. (Smallest deviations in bold.)}
\medskip
\label{tab:cmp}
\scriptsize
\begin{tabular}{lrrrrlrrrlrrrrlrrr}
\hline\noalign{\smallskip}
\multicolumn{1}{l}{\textbf{Group}} & \multicolumn{ 4}{c}{\textbf{CS}} & \multicolumn{1}{c}{\textbf{}} & \multicolumn{3}{c}{\textbf{TS}} & \multicolumn{1}{c}{\textbf{}} & \multicolumn{ 4}{c}{\textbf{HGA}} & \multicolumn{1}{c}{\textbf{}} & \multicolumn{ 3}{c}{\textbf{IBS}} \\
\cline{2-5}\cline{7-9}\cline{11-14}\cline{16-18}
\multicolumn{1}{r}{} & \multicolumn{1}{r}{best} & \multicolumn{1}{r}{avrg} & \multicolumn{1}{r}{t(s)} & \multicolumn{1}{r}{$t_b$ (s)} &  & \multicolumn{1}{r}{best} & \multicolumn{1}{r}{t(s)} & \multicolumn{1}{r}{$t_b$ (s)} &  & \multicolumn{1}{r}{best} & \multicolumn{1}{r}{avrg} & \multicolumn{1}{r}{t(s)} & \multicolumn{1}{r}{$t_b$ (s)} &  & \multicolumn{1}{r}{best} & \multicolumn{1}{r}{avrg} & \multicolumn{1}{r}{$t_b$ (s)} \\
\noalign{\smallskip}\hline\noalign{\smallskip}
Roszieg 1                             & \textbf{0.0\%} & 0.3\%          & 1.8  & 0.2  &  & \textbf{0.0\%} & 2.1  & 0.0 &  & \t.extbf{0.0\%}  & \textbf{0.0\%}  & 3.3   & 0.0   &  & \textbf{0.0\%} & \textbf{0.0\%} & 0.0   \\ 
Roszieg 2                             & \textbf{0.0\%} & 1.3\%          & 1.8  & 0.1  &  & \textbf{0.0\%} & 2.5  & 0.0 &  & \textbf{0.0\%}  & 0.1\%           & 4.5   & 0.0   &  & \textbf{0.0\%} & \textbf{0.0\%} & 0.1   \\ 
Roszieg 3                             & \textbf{0.0\%} & 0.1\%          & 1.8  & 0.1  &  & \textbf{0.0\%} & 1.9  & 0.1 &  & \textbf{0.0\%}  & \textbf{0.0\%}  & 4.0   & 0.0   &  & \textbf{0.0\%} & \textbf{0.0\%} & 0.1   \\ 
Roszieg 4                             & \textbf{0.0\%} & \textbf{0.0\%} & 1.8  & 0.0  &  & \textbf{0.0\%} & 1.9  & 0.0 &  & \textbf{0.0\%}  & \textbf{0.0\%}  & 3.4   & 0.0   &  & \textbf{0.0\%} & \textbf{0.0\%} & 0.0   \\ 
Roszieg 5                             & 2.2\%          & 2.2\%          & 2.4  & 0.0  &  & \textbf{0.0\%} & 2.4  & 0.0 &  & \textbf{0.0\%}  & \textbf{0.0\%}  & 3.6   & 0.0   &  & \textbf{0.0\%} & \textbf{0.0\%} & 0.0   \\ 
Roszieg 6                             & \textbf{0.0\%} & 5.9\%          & 2.4  & 0.3  &  & \textbf{0.0\%} & 2.6  & 0.1 &  & 1.0\%           & 1.1\%           & 4.0   & 0.1   &  & \textbf{0.0\%} & \textbf{0.0\%} & 0.0   \\ 
Roszieg 7                             & \textbf{0.0\%} & 0.5\%          & 2.4  & 0.1  &  & \textbf{0.0\%} & 2.5  & 0.0 &  & \textbf{0.0\%}  & \textbf{0.0\%}  & 4.5   & 0.0   &  & \textbf{0.0\%} & \textbf{0.0\%} & 0.0   \\ 
Roszieg 8                             & 0.6\%          & 1.9\%          & 2.4  & 0.1  &  & \textbf{0.0\%} & 2.6  & 0.0 &  & \textbf{0.0\%}  & \textbf{0.0\%}  & 4.5   & 0.1   &  & \textbf{0.0\%} & \textbf{0.0\%} & 0.0   \\ 
\multicolumn{1}{l}{\textbf{Average}} & 0.4\%          & 1.5\%          & 2.1  & 0.1  &  & \textbf{0.0\%} & 2.3  & 0.0 &  & 0.1\%           & 0.1\%           & 4.0   & 0.0   &  & \textbf{0.0\%} & \textbf{0.0\%} & 0.0   \\ \hline
Heskia 1                              & \textbf{0.0\%} & 0.2\%          & 2.3  & 0.4  &  & \textbf{0.0\%} & 3.1  & 0.4 &  & \textbf{0.0\%}  & \textbf{0.0\%}  & 6.9   & 0.2   &  & \textbf{0.0\%} & \textbf{0.0\%} & 8.2   \\ 
Heskia 2                              & \textbf{0.0\%} & 0.2\%          & 2.3  & 0.2  &  & 0.1\%          & 3.0  & 0.0 &  & 0.1\%           & 0.1\%           & 9.3   & 0.3   &  & \textbf{0.0\%} & \textbf{0.0\%} & 3.0   \\ 
Heskia 3                              & \textbf{0.0\%} & 0.0\%          & 2.3  & 0.2  &  & \textbf{0.0\%} & 3.0  & 0.0 &  & \textbf{0.0\%}  & \textbf{0.0\%}  & 9.2   & 0.3   &  & \textbf{0.0\%} & \textbf{0.0\%} & 5.6   \\ 
Heskia 4                              & \textbf{0.0\%} & 0.1\%          & 2.3  & 0.5  &  & 0.1\%          & 3.0  & 0.1 &  & \textbf{0.0\%}  & 0.3\%           & 9.5   & 0.5   &  & \textbf{0.0\%} & \textbf{0.0\%} & 5.2   \\ 
Heskia 5                              & 0.2\%          & 1.2\%          & 3.3  & 0.3  &  & 1.2\%          & 4.1  & 0.3 &  & \textbf{0.0\%}  & 0.5\%           & 8.0   & 0.2   &  & \textbf{0.0\%} & \textbf{0.0\%} & 1.1   \\ 
Heskia 6                              & 0.5\%          & 2.4\%          & 3.3  & 0.5  &  & 1.5\%          & 4.5  & 0.4 &  & \textbf{0.0\%}  & 0.6\%           & 7.4   & 0.3   &  & \textbf{0.0\%} & \textbf{0.0\%} & 2.5   \\ 
Heskia 7                              & \textbf{0.0\%} & 1.1\%          & 3.3  & 0.1  &  & 0.6\%          & 4.4  & 0.1 &  & \textbf{0.0\%}  & 0.3\%           & 6.6   & 0.2   &  & \textbf{0.0\%} & \textbf{0.0\%} & 1.7   \\ 
Heskia 8                              & 0.2\%          & 1.4\%          & 3.3  & 0.4  &  & 2.3\%          & 4.2  & 0.0 &  & \textbf{0.0\%}  & 0.7\%           & 9.2   & 1.5   &  & \textbf{0.0\%} & \textbf{0.0\%} & 2.5   \\ 
\multicolumn{1}{l}{\textbf{Average}} & 0.1\%          & 0.8\%          & 2.8  & 0.3  &  & 0.7\%          & 3.7  & 0.2 &  & \textbf{0.0\%}  & 0.3\%           & 8.3   & 0.4   &  & \textbf{0.0\%} & \textbf{0.0\%} & 3.7   \\ \hline
Wee-Mag 1                             & 2.5\%          & 8.3\%          & 44.9 & 6.3  &  & 9.4\%          & 46.7 & 0.0 &  & \textbf{-6.3\%} & \textbf{-2.4\%} & 136.9 & 56.8  &  & 0.6\%          & 4.3\%          & 104.9 \\ 
Wee-Mag 2                             & 6.0\%          & 11.6\%         & 44.8 & 9.9  &  & 14.7\%         & 47.3 & 0.0 &  & \textbf{-3.6\%} & \textbf{0.4\%}  & 158.8 & 60.1  &  & 0.6\%          & 4.5\%          & 89.8  \\ 
Wee-Mag 3                             & 3.9\%          & 12.5\%         & 44.5 & 17.4 &  & 17.2\%         & 48.1 & 3.9 &  & \textbf{-3.2\%} & \textbf{1.4\%}  & 248.5 & 115.8 &  & 1.8\%          & 4.8\%          & 161.5 \\ 
Wee-Mag 4                             & 1.5\%          & 10.0\%         & 44.6 & 16.3 &  & 14.6\%         & 49.5 & 0.0 &  & \textbf{-5.8\%} & \textbf{-1.4\%} & 245.9 & 112.7 &  & 0.4\%          & 4.2\%          & 135.5 \\ 
Wee-Mag 5                             & 20.4\%         & 20.4\%         & 60.0 & 1.2  &  & 20.4\%         & 56.4 & 0.0 &  & 2.2\%           & 7.9\%           & 213.9 & 61.4  &  & \textbf{0.0\%} & \textbf{3.7\%} & 54.1  \\ 
Wee-Mag 6                             & 18.8\%         & 20.5\%         & 60.4 & 1.6  &  & 20.6\%         & 57.6 & 0.0 &  & 3.7\%           & 8.2\%           & 225.6 & 66.1  &  & \textbf{0.0\%} & \textbf{3.9\%} & 59.3  \\ 
Wee-Mag 7                             & 13.7\%         & 19.9\%         & 59.2 & 5.4  &  & 20.8\%         & 57.3 & 0.0 &  & 2.4\%           & 8.2\%           & 283.7 & 97.9  &  & \textbf{0.0\%} & \textbf{4.3\%} & 77.6  \\ 
Wee-Mag 8                             & 15.3\%         & 19.7\%         & 59.2 & 3.9  &  & 20.2\%         & 56.1 & 0.0 &  & 1.9\%           & 6.7\%           & 288.1 & 108.9 &  & \textbf{0.0\%} & \textbf{4.2\%} & 90.0  \\ 
\multicolumn{1}{l}{\textbf{Average}} & 10.3\%         & 15.4\%         & 52.2 & 7.7  &  & 17.2\%         & 52.4 & 0.5 &  & \textbf{-1.1\%} & \textbf{3.6\%}  & 225.2 & 84.9  &  & 0.4\%          & 4.2\%          & 96.6  \\ \hline
Tonge 1                               & 7.3\%          & 9.6\%          & 45.9 & 2.5  &  & 7.9\%          & 32.2 & 4.8 &  & \textbf{-1.3\%} & \textbf{2.1\%}  & 205.7 & 34.4  &  & 1.0\%          & 2.8\%          & 86.4  \\ 
Tonge 2                               & 8.1\%          & 12.0\%         & 46.5 & 2.9  &  & 9.9\%          & 33.5 & 3.5 &  & \textbf{-0.8\%} & \textbf{0.9\%}  & 241.2 & 34.9  &  & 0.0\%          & 1.2\%          & 92.2  \\ 
Tonge 3                               & 6.9\%          & 11.5\%         & 46.3 & 5.6  &  & 8.8\%          & 34.5 & 6.7 &  & \textbf{-1.0\%} & \textbf{1.4\%}  & 391.0 & 98.6  &  & 0.6\%          & 2.5\%          & 160.1 \\ 
Tonge 4                               & 5.2\%          & 11.4\%         & 46.3 & 4.2  &  & 7.8\%          & 32.3 & 7.0 &  & \textbf{-0.3\%} & \textbf{1.2\%}  & 347.6 & 56.9  &  & 0.6\%          & 1.5\%          & 171.4 \\ 
Tonge 5                               & 15.8\%         & 15.8\%         & 48.4 & 1.2  &  & 15.5\%         & 41.1 & 0.0 &  & 3.0\%           & 6.2\%           & 296.9 & 74.0  &  & \textbf{0.0\%} & \textbf{3.3\%} & 88.0  \\ 
Tonge 6                               & 13.9\%         & 13.9\%         & 48.8 & 1.2  &  & 12.4\%         & 40.8 & 0.0 &  & 0.6\%           & 5.1\%           & 300.0 & 67.1  &  & \textbf{0.0\%} & \textbf{2.5\%} & 70.5  \\ 
Tonge 7                               & 16.0\%         & 16.5\%         & 49.0 & 1.3  &  & 16.5\%         & 40.5 & 0.0 &  & 0.4\%           & 5.1\%           & 446.7 & 129.1 &  & \textbf{0.0\%} & \textbf{2.3\%} & 124.3 \\ 
Tonge 8                               & 17.1\%         & 17.5\%         & 49.0 & 1.3  &  & 14.8\%         & 39.6 & 7.1 &  & 1.7\%           & 4.3\%           & 469.4 & 105.3 &  & \textbf{0.0\%} & \textbf{2.9\%} & 156.4 \\ 
\multicolumn{1}{l}{\textbf{Average}} & 11.3\% & 13.5\% & 47.5 & 2.5 & & 11.7\% & 36.8 & 3.6 &  &  \textbf{0.3\%}  & 3.3\%  & 337.3 & 75.1 &  & \textbf{0.3\%} & \textbf{2.4\%} & 118.7 \\ 
\hline
\multicolumn{1}{l}{\textbf{Overall}} & 5.7\% & 7.8\% & \multicolumn{1}{l}{} & \multicolumn{1}{l}{} &  & 7.4\% & \multicolumn{1}{l}{} & \multicolumn{1}{l}{} &  & \textbf{-0.2\%} & 1.8\% & \multicolumn{1}{l}{} & \multicolumn{1}{l}{} &  & \textbf{0.2\%} & \textbf{1.7\%} & \multicolumn{1}{l}{} \\ 
\end{tabular}
\end{sidewaystable}

All meta-heuristics have similar computation times. The results of the
CS have been obtained on a PC with a Pentium $4$ $2.6$ GHz with $1$ GB
main memory, the TS results on a PC with a $1.66$ GHz Intel Core 2 Duo
T5450 with $3$ GB main memory, and the results of the IBS on a PC with
a $2.6$ GHz Pentium $4$ processor and $1$ GB of main memory. The
computation time on these architectures may vary about a factor of
two, but we consider the total computation time of all methods to be
reasonable for an NP-hard optimization problem.

Concerning the solution quality, for the smaller instance families
Heskia and Roszieg, all methods obtain good results, with a relative
deviation of the average value of at most $1.5$ percent (CS) and a
relative deviation of the best value of at most $0.4$ percent
(CS). IBS and TS solve these instances optimally. The small relative
deviations of the HGA come from two of the $80$ instances, which were
not solved to optimality. On the larger instance families Wee-Mag and
Tonge, IBS and the HGA always outperform CS and TS, with average
relative deviations in each group of instances at least four percent
better. The HGA obtains better best values in eight of the $16$ groups
corresponding to $73$ of the $160$ instances. The HGA is able to
improve the average values in eight of the $16$
groups. Table~\ref{tab:compibshga} summarizes the number of instances
in which IBS and HGA obtain a better average and best value.

\begin{table}
  \centering
  \caption{Comparison of IBS and HGA.}
  \begin{tabular}{lccc}
    \hline
    & IBS better & Ties & HGA better\\
    \hline
    Avg. & 93 & 143 & 84\\
    Best & 48 & 199 & 73\\
    \hline
  \end{tabular}
  \label{tab:compibshga}
\end{table}

\begin{figure}
  \centering
  \includegraphics[width=0.99\textwidth]{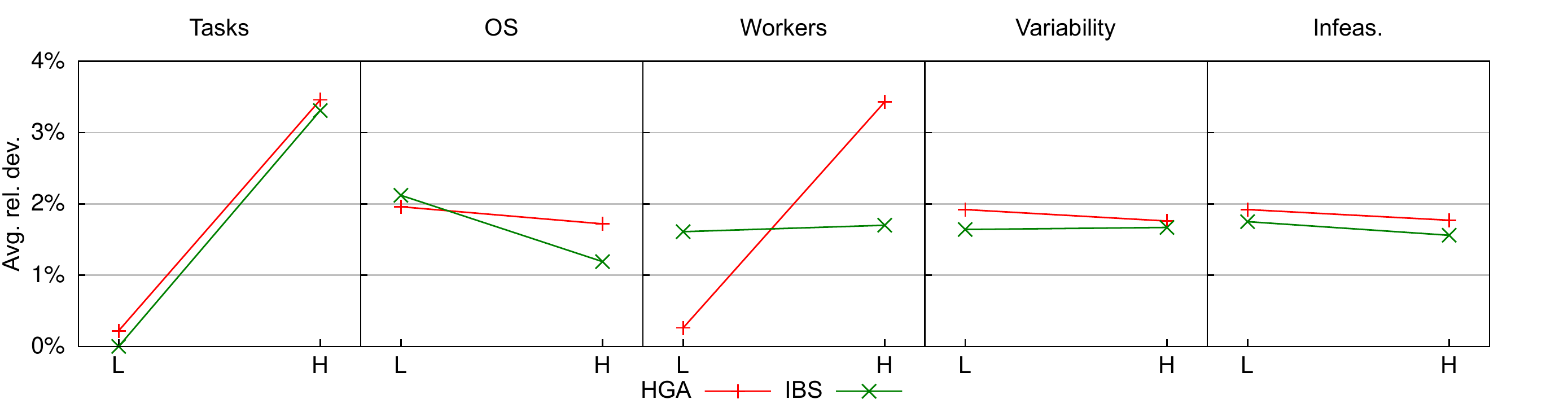}\\\includegraphics[width=0.99\textwidth]{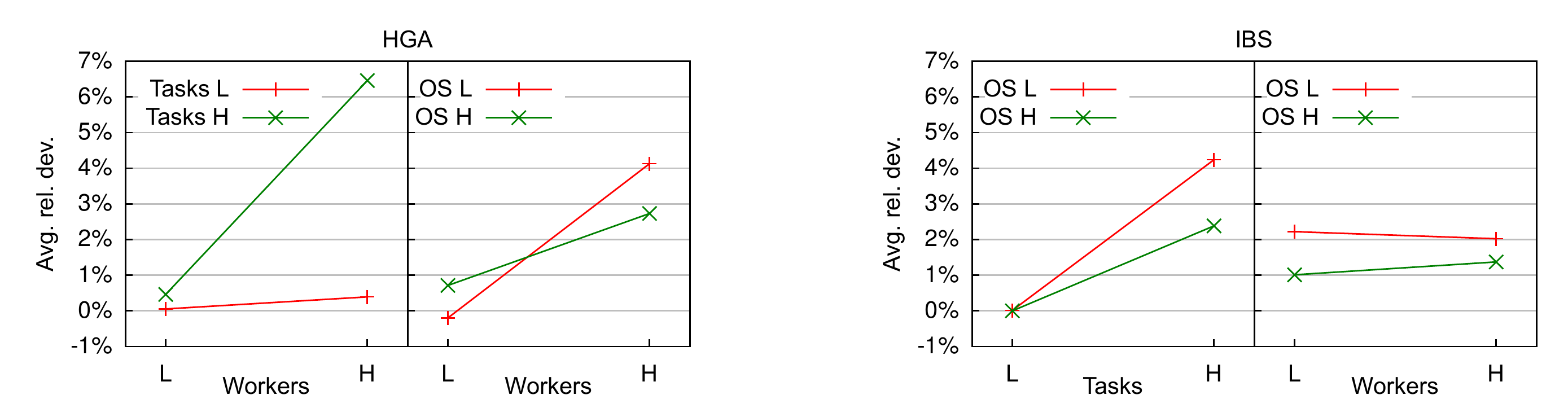}
  \caption{Main effects (upper row) and most significant double
    interactions (lower row) for HGA and IBS of the five experimental
    factors on average relative deviations.}
  \label{fig:interaction}
\end{figure}

A statistical analysis shows that the most significant factors
influencing the average solution quality of HGA and IBS are the number
of tasks and workers, and the order
strength. Figure~\ref{fig:interaction} shows the main effects and
interaction plots for the two most significant double interactions for
HGA and IBS.  The solution quality of both methods strongly depends on
the number of tasks. IBS performs better for low order strength, and
the HGA performs significantly better when the number of workers is
low. These dependencies are more pronounced for a high number of
tasks.

In summary, the approach using a HGA seems adequate for obtaining good
solutions, with results comparable to or better than the existing
heuristics. In particular it obtains significantly better solutions
for a low number of workers. Being a relatively simple method,
optimizing the task priorities of a station-based assignment scheme
inside a generic hybrid genetic algorithm, the HGA may be an
interesting alternative for solving the ALWABP-2.

\section{Conclusions}
\label{sec:conclusions}

In this paper, we have proposed a constructive heuristic framework
based on task and worker assignment priority rules for the assembly
line worker assignment and balancing problem. In a series of
computational tests, the approach proved to be fast and efficient both
when evaluated as a stand-alone method as well as when the obtained
solutions were used to improve the convergence of more elaborate
meta-heuristics. Moreover, the strategy was used as a solution decoder
within a hybrid genetic algorithm which was also proposed and tested,
obtaining results that were comparable to the best known methods
available in the literature.

\section*{Acknowledgements}
This research was supported by the Brazilian Conselho Nacional de
Desenvolvimento Científico e Tecnológico (CNPq, Brazil) and by
Fundação de Amparo à Pesquisa do Estado de São Paulo (FAPESP,
Brazil). This support is gratefully acknowledged.

\bibliographystyle{spbasic}
\bibliography{abbreviations,marcus,library}

\begin{thebibliography}{19}
\providecommand{\natexlab}[1]{#1}
\providecommand{\url}[1]{{#1}}
\providecommand{\urlprefix}{URL }
\expandafter\ifx\csname urlstyle\endcsname\relax
  \providecommand{\doi}[1]{DOI~\discretionary{}{}{}#1}\else
  \providecommand{\doi}{DOI~\discretionary{}{}{}\begingroup
  \urlstyle{rm}\Url}\fi
\providecommand{\eprint}[2][]{\url{#2}}

\bibitem[{Baybars(1986)}]{baybars86survey}
Baybars I (1986) A survey of exact algorithms for the simple assembly line
  balancing problem. Manag Sci 32:909--932

\bibitem[{Bean(1994)}]{Bean/1994}
Bean JC (1994) Genetic algorithms and random keys for sequencing and
  optimization. ORSA J Comp 6:154--160

\bibitem[{Blum and Miralles(2011)}]{blum11solving}
Blum C, Miralles C (2011) On solving the assembly line worker assignment and
  balancing problem via beam search. Comp Oper Res 38:328--339

\bibitem[{Boysen et~al(2007)Boysen, Fliedner, and
  Scholl}]{boysen07classification}
Boysen N, Fliedner M, Scholl A (2007) A classification of assembly line
  balancing problems. Eur J Oper Res 183:674--693,
  \doi{10.1016/j.ejor.2006.10.010}

\bibitem[{Boysen et~al(2008)Boysen, Fliedner, and Scholl}]{boysen08assembly}
Boysen N, Fliedner M, Scholl A (2008) Assembly line balancing: Which model to
  use when? Int J Prod Econ 111:509--528, \doi{10.1016/j.ijpe.2007.02.026}

\bibitem[{Chaves et~al(2007)Chaves, Miralles, and Lorena}]{chaves07clustering}
Chaves AA, Miralles C, Lorena LAN (2007) Clustering search approach for the
  assembly line worker assignment and balancing problem. In: Proc. of the 37th
  International Conference on Computers and Industrial Engineering, Alexandria,
  Egypt, pp 1469--1478

\bibitem[{Chaves et~al(2009)Chaves, Lorena, and Miralles}]{chaves09hybridb}
Chaves AA, Lorena LAN, Miralles C (2009) Hybrid metaheuristic for the assembly
  line worker assignment and balancing problem. Lecture Notes on Computer
  Science 5818:1--14

\bibitem[{Goldberg(1989)}]{Goldberg/1989}
Goldberg DE (1989) Genetic Algorithms in Search, Optimization, and Machine
  Learning. Addison-Wesley

\bibitem[{{Gon\c{c}alves} and de~Almeida(2004)}]{Goncalves.Almeida/2004}
{Gon\c{c}alves} JF, de~Almeida JR (2004) A hybrid genetic algorithm for
  assembly line balancing. J Heuristics 8(6):629--642

\bibitem[{Holland(1975)}]{Holland.1975}
Holland JH (1975) Adaptation in Natural and Artifical Systems. MIT Press,
  Cambrigde, MA.

\bibitem[{Klein and Scholl(1999)}]{Klein.Scholl/1999}
Klein R, Scholl A (1999) Computing lower bounds by destructive improvement --
  an application to resource-constrained project scheduling. Eur J Oper Res
  112:322--346

\bibitem[{Mendes et~al(2009)Mendes, {Gon\c{c}alves}, and
  Resende}]{Mendes.etal/2009}
Mendes JJM, {Gon\c{c}alves} JF, Resende MGC (2009) A random key based genetic
  algorithm for the resource constrained project scheduling problem. Comp Oper
  Res 36(1):92--109

\bibitem[{Miralles et~al(2007)Miralles, Garcia-Sabater, Andr\'{e}s, and
  Cardos}]{miralles07advantages}
Miralles C, Garcia-Sabater JP, Andr\'{e}s C, Cardos M (2007) Advantages of
  assembly lines in sheltered work centres for disabled. {A} case study. Int J
  Prod Econ 110:187--197

\bibitem[{Miralles et~al(2008)Miralles, Garcia-Sabater, Andr\'{e}s, and
  Cardos}]{miralles08branch}
Miralles C, Garcia-Sabater JP, Andr\'{e}s C, Cardos M (2008) Branch and bound
  procedures for solving the assembly line worker assignment and balancing
  problem: Application to sheltered work centres for disabled. Discrete App
  Math 156:352--367

\bibitem[{Moreira and Costa(2009)}]{moreira09minimalist}
Moreira MCO, Costa AM (2009) A minimalist yet efficient tabu search for
  balancing assembly lines with disabled workers. In: Anais do XLI Simp\'{o}sio
  Brasileiro de Pesquisa Operacional, Porto Seguro

\bibitem[{Scholl(1999)}]{scholl99balancing}
Scholl A (1999) Balancing and sequencing of assembly lines. Physica-Verlag

\bibitem[{Scholl and Becker(2006)}]{Scholl.Becker/2006}
Scholl A, Becker C (2006) State-of-the-art exact and heuristic solution
  procedures for simple assembly line balancing. Eur J Oper Res 168(3):666--693

\bibitem[{Scholl and {Vo\ss}(1996)}]{scholl96simple}
Scholl A, {Vo\ss} S (1996) Simple assembly line balancing -- heuristic
  approaches. J Heuristics 2:217--244, \doi{10.1007/BF00127358}

\bibitem[{Talbot et~al(1986)Talbot, Patterson, and Gehrlein}]{Talbot.etal/1986}
Talbot FB, Patterson JH, Gehrlein WV (1986) A comparative evaluation of
  heuristic line balancing techniques. Manag Sci 32(4):430--454

\end{thebibliography}

\end{document}